\documentclass{sigkddExp}

\begin{document}

\title{A Baseline for Content-Based Blog Classification}

\numberofauthors{2}

\author{
\alignauthor Olof G\"{o}rnerup\\
       \affaddr{Swedish Institute of Computer Science (SICS)}\\
       \affaddr{Box 1263}\\
       \affaddr{SE-16429 Kista, Sweden}\\
       \email{olofg@sics.se}
\alignauthor Magnus Boman\\
       \affaddr{Swedish Institute of Computer Science (SICS)}\\
       \affaddr{Box 1263}\\
       \affaddr{SE-16429 Kista, Sweden}\\
       \email{mab@sics.se}
}

\date{29 July 2099}
\maketitle
\begin{abstract}

A content-based network representation of web logs (blogs) using a basic word-overlap similarity measure is presented. Due to a strong signal in blog data the approach is sufficient for accurately classifying blogs. Using Swedish blog data we demonstrate that blogs that treat similar subjects are organized in clusters that, in turn, are hierarchically organized in higher-order clusters. The simplicity of the representation renders it both computationally tractable and transparent. We therefore argue that the approach is suitable as a baseline when developing and analyzing more advanced content-based representations of the blogosphere.

\end{abstract}

\section{Introduction}
For any given blog, a number of other blogs is related to it through content overlap. The owner of the blog and its readers---or indeed anyone interested in blog navigation in general---are probably interested in learning more about those related blogs.
The problem is that the sheer size of the blogosphere makes keeping up with the dynamic set of related blogs next to impossible.
To assist, a vast array of tools and algorithms have been presented (cf. \cite{agarwal:blogosphere,AgarwalLiu09}). While it is important
to remember that blog networks differ from general Web page networks in several important respects \cite{agarwal:blogosphere},
much of the mathematics governing the latter is reusable for modeling also the former (see, e.g., \cite{kleinberg:authoritative}). The classification of blogs can be made according to different criteria, including blog entry similarity \cite{tauro:vizblog}, 
interblog communication and community stability
 \cite{chi:structural}, sense of community among bloggers \cite{chin:social}, 
discussion keyword correlation \cite{bansal:seeking}, and a host of machine learning and statistics approaches. To date, however, almost all these tools and algorithms
require human intervention and considerable time investment to overcome problems with bootstrapping, tuning, and not least
semantics. Understanding a graph, perhaps with thousands of vertices and edges, pertaining to describe relevance to one's own blog 
according to some set of possibly esoteric or advanced criteria is not straightforward. We address this problem by presenting
a method for generating a network of relevant blogs by means of the simplest similarity criterion there is: word overlap. We will demonstrate that even this na\"{i}ve approach allows us to accurately classify blogs with respect to their contents, filter out spam weblogs (splogs) and multiple occurrences (i.e., blogs
linked to from multiple URLs) in addition to providing a global overview of the blogosphere. The procedure is transparent, modular, computationally efficient,
and requires no human monitoring or control. Using a network representation also enables one to employ a wealth of theory and techniques recently developed in network theory \cite{Newman2006}.

We describe our methodology in the section that follows. Section 3 presents results of our experiments, using data from the Swedish blogosphere. We conclude the paper by discussing the results and pointing to possible future directions in Section 4.

\section{Methodology}
Our overall approach is to provide a global view of the blogosphere in the form of a network, where vertices constitute blogs and where weighted edges constitute similarity relations between blogs. There is a link between two blogs if they are related, and the strength of the link is given by the degree of similarity. 

\subsection{Similarity measure}

To estimate the similarity between blogs we simply compare the overlap of occurring words. In more formal terms, the similarity is quantified in the following way. 
Given two blogs $i$ and $j$, let $\mathcal{W}_i$ denote a set of words (to be qualified below) that occur in $i$ and $\mathcal{W}_j$ a set of words that are used in $j$. 
The similarity $s_{ij}$ between $i$ and $j$ is then defined as
\begin{equation}
s_{ij}=\frac{|\mathcal{W}_i \cap \mathcal{W}_j|}{ |\mathcal{W}_i \cup \mathcal{W}_j|}.
\label{sim_eq}
\end{equation}
In other words, $s_{ij}$ is the fraction of all words in $\mathcal{W}_i$ and $\mathcal{W}_j$ that are shared by the two sets. It holds that $0 \leq s_{ij} \leq 1$, where $s_{ij}=1$ if $\mathcal{W}_i$ and $\mathcal{W}_j$ are identical and $s_{ij}=0$ if they do not share a single word. The similarity measure is equivalent to Tversky's Ratio model \cite{tversky1977}, which has been found to be a good trade-off between simplicity and performance among text document similarity models \cite{lee2005}.

We do not consider the full word sets of blogs---literally \emph{all} occurring words---for several reasons. Firstly, comparing very common words (``the'', ``it'', ``do'', etc.) will only provide a negligible amount of similarity information. The use of very uncommon words, on the other hand, is likely to tell us a lot about the characteristics of a blog. However, at the same time we do not want to consider words that are \emph{too} uncommon---for instance occurring only a handful of times in the blogosphere during the course of several months---since these are often misspellings and typos that only add noise to the statistics. Another reason for not considering all words is a pragmatic one. Analyzing tens of thousands of blogs can be computationally expensive. By utilizing Zipf's law \cite{Zipf49}, which implies that a few of the most common words stand for a large majority of word occurrences\footnote{More specifically, the frequency of a word is inversely proportional to its rank; $f_n \sim 1/n^a$, where $n$ is the rank ($n=1$ for the most common word, $n=2$ for the second most common word, etc.) and $a$ is some exponent.}, the computational cost is drastically reduced. 

\subsection{Network structure}
The global structure of a similarity network provides valuable information that facilitates an exploration of the blogosphere. Specifically, blogs are organized in clusters that reflect domains of topics such as politics, books, technology, or music. What characterizes clusters is that there are significantly higher densities of edges within clusters than between them. Such clusters are referred to as \emph{communities} in the network literature. We chose to use the term cluster here, however, due to the ambiguity of the word ``community'' in this context. A cluster may indeed constitute a community in the social meaning of the word, although it is not necessarily so. In fact, one advantage of using a content-based similarity measure is that it relates blogs that otherwise lack explicit (social or hyper-) links.

Clusters can be quantified as follows \cite{Newman04}. Let $\{v_1,v_2,..., v_n\}$ be a partition of a set of vertices into $n$ groups, $r_{i}$ the degree of edge weights (i.e., similarities) internal to $v_i$ (the sum of internal weights over the sum of all weights in the network) and $s_i$ the degree of weights of edges that start in  $v_i$. The degree of cluster structure is then defined as
\begin{equation}
Q = \sum_{i=1}^n (r_{i}-s_i^2).
\end{equation}
To infer clusters in the blog network we employ Clauset's method \cite{Clauset05}, which aims to find cluster assignments---a partition of the set of vertices---such that $Q$ is maximized.

Edges are not only organized in clusters, but the clusters are in turn organized in higher order clusters. This hierarchical organization of the graph is interesting since it may enable automatic generation of hierarchical blog taxonomies. Here we use a Monte Carlo sampling technique by Clauset et al.~\cite{clauset08} to identify hierarchical structure in the blog network.

\subsection{Case study: The Swedish blogosphere}

We have tested our approach on the Swedish blogosphere. The API of the blog search engine \emph{Twingly}\footnote{http://www.twingly.com/} was used to collect data, where individual blog posts from the period March-July 2009 were fetched and indexed. In the spirit of keeping things simple, we refrained from applying ad hoc textbook pre-processing, such as stemming words or removing slang or acronyms (with respect to predefined word lists) \cite{AgarwalLiu09}. Instead we relied on basic word frequency statistics to filter out words in the following specific way. During indexing, word frequencies were calculated. We then discarded all words occurring less than ten times. Of the remaining words we kept those that occured in the fifth percentile of the frequency distribution. For each blog, we collected its set of those occurring words. Blogs that had word sets of size 25 or larger were kept. This ensured a meaningful similarity measure and also filtered out a considerable amount of spam blogs. At this point, 21564 blogs remained. The similarities between each pair of these were calculated, and stored if equal to or above 0.025. We have tested to vary the above parameters and the results reported here appear to be stable.

\section{Results}

\begin{figure}
\begin{center}
\includegraphics[scale=1.0]{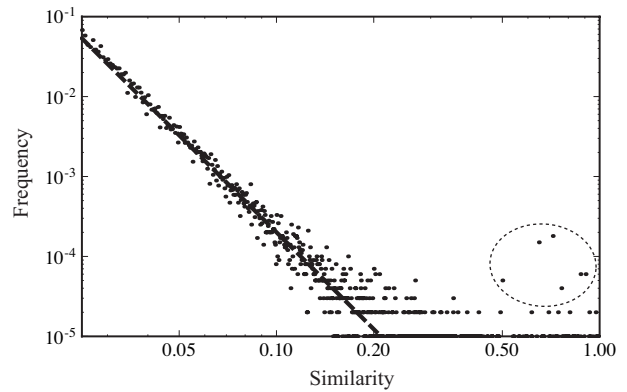}
\caption{Log-log plot of the similarity distribution of blogs. The dashed line denotes a least-squares linear regression in the region $[0.025, 0.2]$ which has a slope of about $-4.0$. Despite the simplicity of the similarity measure, it provides valuable information: Similarities between blogs that occur in multiple copies (or nearly copies) appear in the very end of the tail, and splogs are easily identified by outliers (enclosed by a dashed ellipse).}
\label{sim_dist}
\end{center}	
\end{figure}

\begin{figure*}
\begin{center}
\includegraphics[scale=1.05]{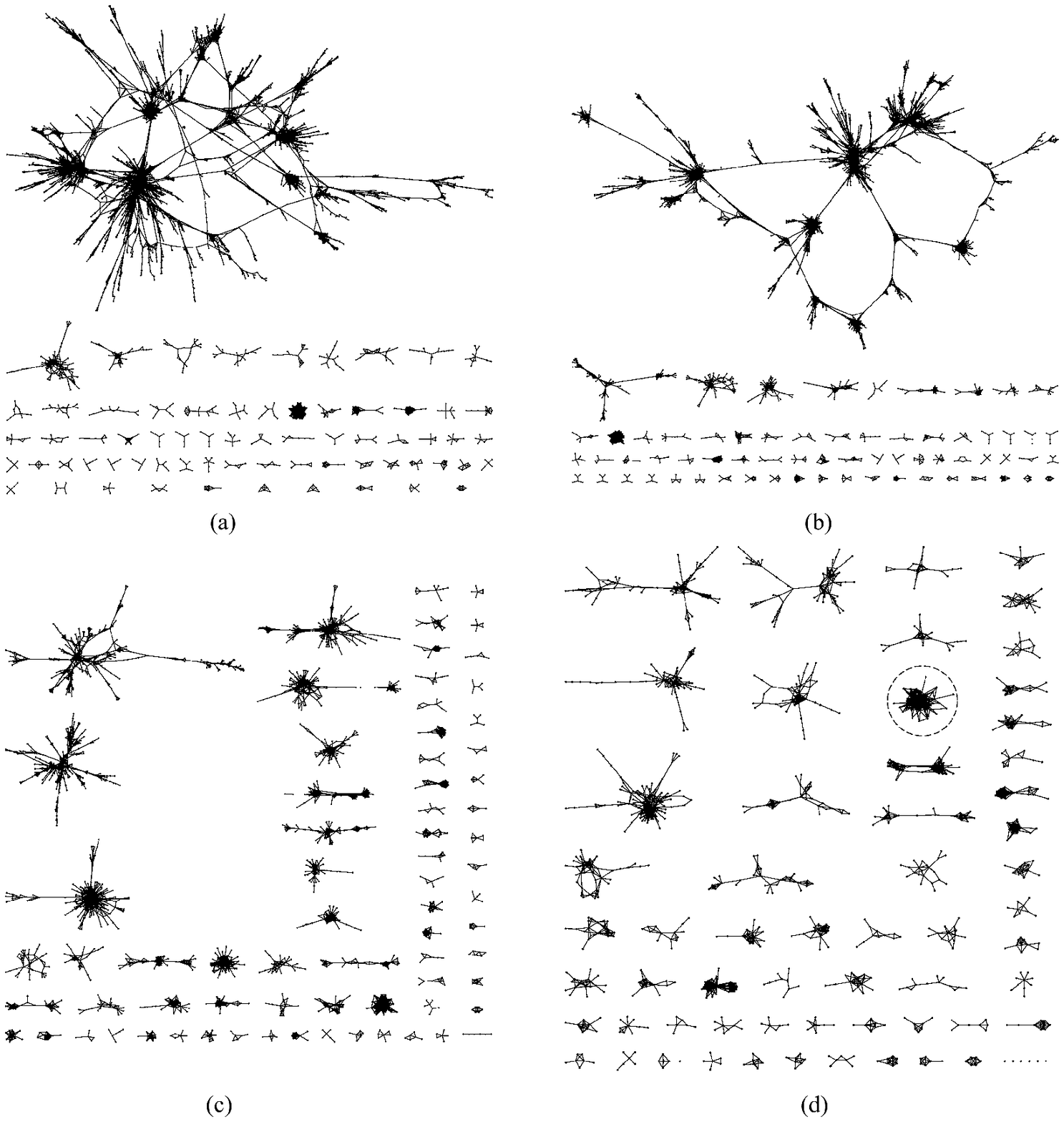}
\caption{Visualization of the Swedish blogosphere, where blogs with similarities $\geq \gamma$ are shown. (a) $\gamma=0.04$. (b) $\gamma=0.045$. (c) $\gamma=0.055$. (d) $\gamma=0.07$. A spam blog cluster is enclosed within a dashed circle.}
\label{diff_thresholds}
\end{center}
\end{figure*}

\begin{figure*}
\begin{center}
\includegraphics[scale=1.0]{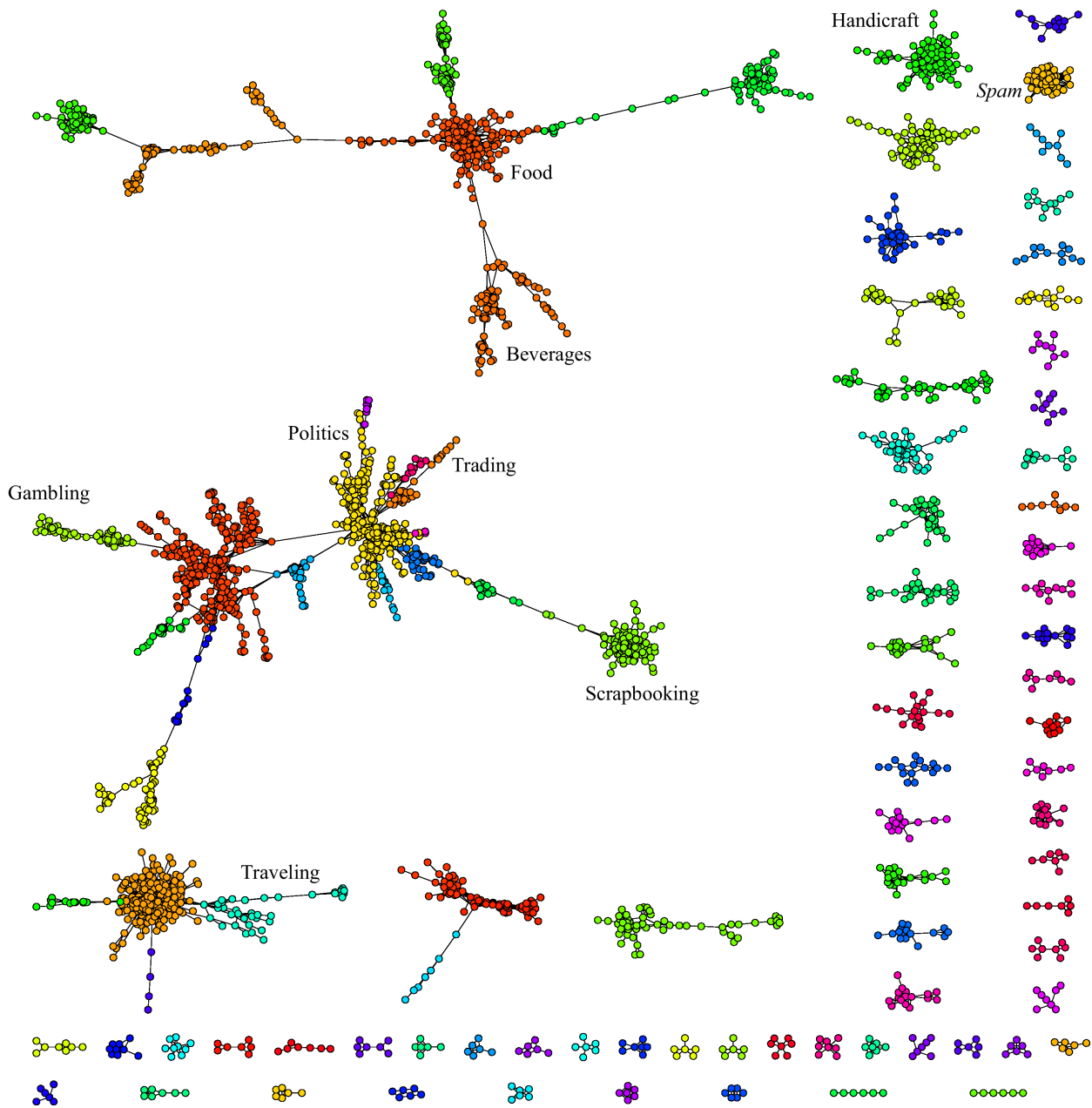}
\caption{Inferred categories of Swedish blogs ($\gamma=0.05$). Some examples are marked with text.}
\label{clusster_net}
\end{center}
\end{figure*}

\begin{figure*}
\begin{center}
\includegraphics[scale=1.6]{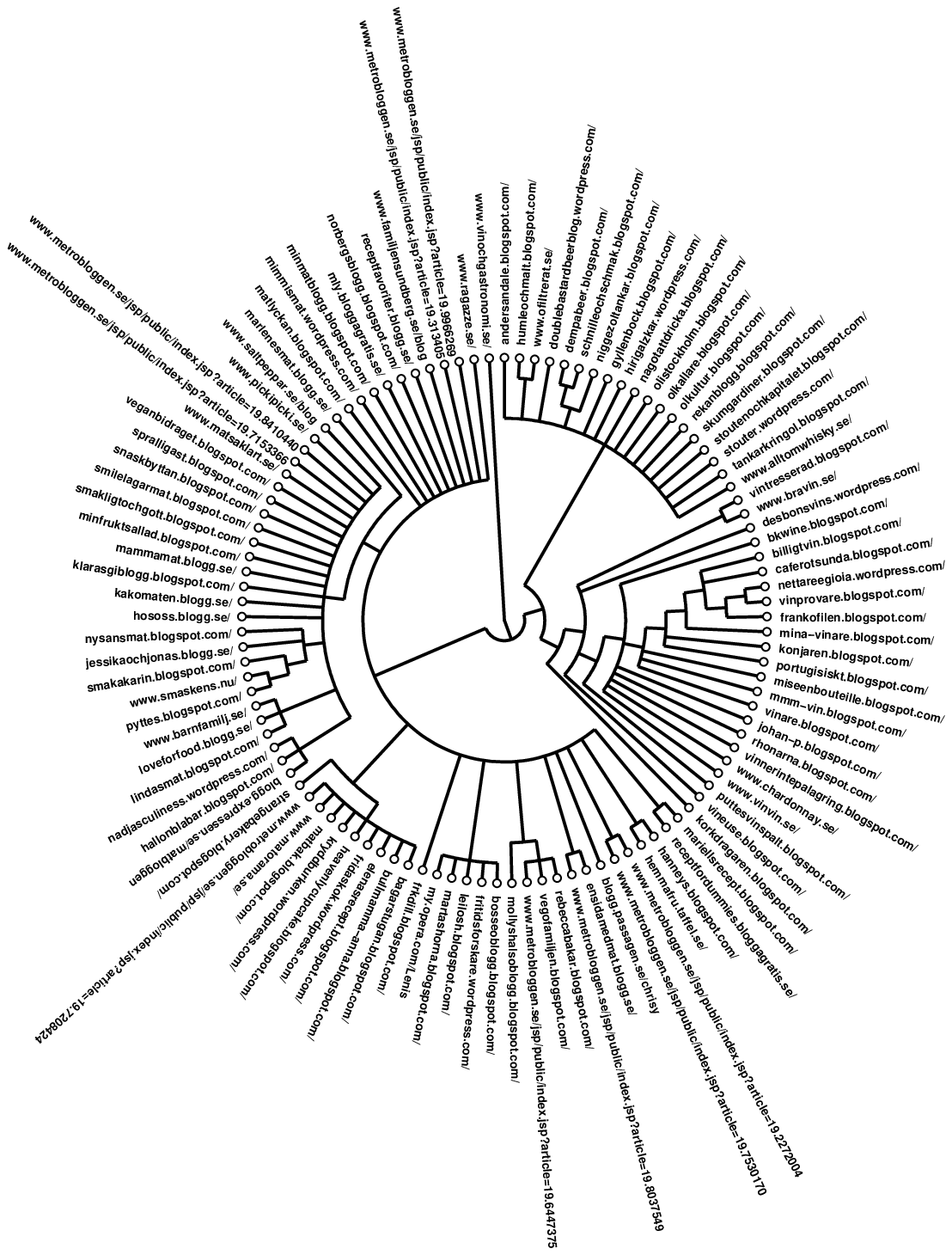}
\caption{Dendrogram of the food-and-beverages cluster. Blogs are in turn organized in separate vine, beer and food clusters. Leaf nodes are labeled with corresponding blog URLs.}
\label{dendro_fig}
\end{center}
\end{figure*}

The distribution of similarities is shown in Fig.~\ref{sim_dist}. Interestingly, remaining splogs are revealed by outliers; similarities that are ``unusually'' high. Groups of multiple occurrences of the same (or slightly deviating) blogs are also easily identified in the tail of the distribution. The acquired network (Fig.~\ref{diff_thresholds}) displays a distinct clustered structure. Splogs are also revealed here by a tightly connected cluster. Fig.~\ref{clusster_net} shows automatically inferred clusters when similarities  $\gamma=0.05$ are considered. When manually sampling and classifying blogs, clusters and blog classes are found to be consistent. An example of the hierarchical organization of the blog network is depicted in Fig.~\ref{dendro_fig} in the form of a dendrogram of a ``food and beverages'' cluster.

\section{Discussion and outlook}

We have shown that the signal in raw blog data is so strong that even our basic similarity measure---word occurrence overlap---is capable of capturing valuable structural information. Since the measure is computationally tractable it therefore enables an efficient mean for classifying blogs when used in concurrence with fast graph clustering algorithms. We grant that there are more advanced---and possibly more accurate---(document) similarity measures, e.g., Latent semantic analysis \cite{Deerwester90} and Random indexing \cite{Kanerva00}. However, we believe that the minimal (non-trivial) measure employed here is suitable as a baseline when studying blog similarity networks. The measure is admittedly simplistic, yet this is also its strength since it decreases the risk of causing hidden representation-dependent artifacts that are more difficult to identify when using more advanced similarity measures.

An issue that needs to be addressed in future work is that of validation. How can we know that acquired blog classifications are correct? So far, our approach has been to examine a random sample of blogs and subjectively confirm that their contents is consistent within inferred blog clusters. Such empirical evaluations can be problematic, however. In some cases, a manual classification is clear cut (i.e.~identifying that two blogs that solely treat Belgian beer belong to the same class), but not always. A more quantitative measure that verifies the result is therefore desirable. This can also be turned into an epistemological question. One can for example imagine cases when the blog classes acquired from the similarity network can be used to evaluate \emph{other} blog classifications (including our own subjective one). However, in this discussion we have more pragmatic and application-oriented evaluation methods in mind.

We have treated only a few structural aspects of the blog network here. These deserve more attention, as well as network dynamics and evolution: How does information diffuse and change in the network, and how does the network structure itself change over time? For instance, through an analysis along these lines one may perhaps trace how emerging trends or news proliferate in and between specific topic domains of the blog similarity network.

Another possible future direction that we have only touched on briefly here concerns splog detection. We have seen that blog classification also groups together splogs. If an individual blog is identified as a splog (e.g., by examining the distribution of blog similarities), it is likely that its associated blog cluster also consists of splogs. If such a relation proves to hold true in general, it enables splog detection and removal at the level of blog clusters rather than individual blogs, which presumably would be much more efficient. 

\subsection*{Acknowledgements}
OG was funded by The Internet Infrastructure Foundation (.SE). The authors thank Twingly for providing blog data and Aaron Clauset for sharing source code for the hierarchical structure inference algorithm and for the radial dendrogram visualization script used for rendering Fig. \ref{dendro_fig}.


\end{document}